\documentclass[journal,comsoc]{IEEEtran}

\usepackage[T1]{fontenc} 
\usepackage{cite}

\usepackage[hidelinks]{hyperref}

\ifCLASSINFOpdf
    \usepackage[pdftex]{graphicx}
    \usepackage{subfig}
    \DeclareGraphicsExtensions{.pdf,.jpeg,.png}
\else
\fi
\usepackage{amsmath, amssymb, amsfonts}

\usepackage[cmintegrals]{newtxmath}
\usepackage{array}
\usepackage{longtable}
\usepackage{rotating}

\usepackage{stfloats}
\usepackage{url}

\usepackage{multirow}
\usepackage{array}
\usepackage{booktabs}

\usepackage{multirow}
\usepackage{enumerate}
\usepackage[linesnumbered,ruled,algo2e] {algorithm2e}
\usepackage{algorithm}
\usepackage{algorithmic}
\usepackage{verbatim}

\usepackage{xargs}
\usepackage[pdftex,dvipsnames]{xcolor}
\usepackage[colorinlistoftodos,prependcaption,textsize=tiny]{todonotes}
\usepackage[section]{placeins}
\newcommandx{\unsure}[2][1=]{\todo[disable,linecolor=red,backgroundcolor=red!25,bordercolor=red,#1]{#2}}
\newcommandx{\change}[2][1=]{\todo[disable,linecolor=blue,backgroundcolor=blue!25,bordercolor=blue,#1]{#2}}
\newcommandx{\info}[2][1=]{\todo[disable,linecolor=OliveGreen,backgroundcolor=OliveGreen!25,bordercolor=OliveGreen,#1]{#2}}
\newcommandx{\improvement}[2][1=]{\todo[disable,linecolor=Plum,backgroundcolor=Plum!25,bordercolor=Plum,#1]{#2}}

\begin{document}

\title{Indoor Fusion Positioning Based on "IMU-Ultrasonic-UWB" and Factor Graph Optimization Method
\thanks{(Corresponding author: Shukai Duan and Shuang-Hua Yang.)}
\thanks{Fengyun Zhang and Shukai Duan are with the College of Artificial Intelligence, Southwest University, Chongqing, China (email: zhangfy24@swu.edu.cn; duansk@swu.edu.cn); Chongqing College of Traditional Chinese Medicine, Chongqing, China (lijia@cqctcm.edu.cn); Xiaoqing Zhang is with the Shenzhen Institute of Advanced Technology, Chinese Academy of Sciences, Shenzhen, China (email: xq.zhang2@siat.ac.cn); Shuang-Hua Yang is with the Department of Computer Science, University of Reading, UK (shuang-hua.yang@reading.ac.uk).}
}

\author{Fengyun~Zhang,~\IEEEmembership{Member,~IEEE,} Jia Li, Xiaoqing Zhang, Shukai Duan, and Shuang-Hua Yang~\IEEEmembership{Senior Member,~IEEE}}

\markboth{IEEE Transactions on Wireless Communications, ~Vol.~xx, No.~x, Mar~2025}%
{}



\maketitle

\begin{abstract}
This paper presents a high-precision positioning system that integrates ultra-wideband (UWB) time difference of arrival (TDoA) measurements, inertial measurement unit (IMU) data, and ultrasonic sensors through factor graph optimization. To overcome the shortcomings of standalone UWB systems in non-line-of-sight (NLOS) scenarios and the inherent drift associated with inertial navigation, we developed a novel hybrid fusion framework. First, a dynamic covariance estimation mechanism is incorporated, which automatically adjusts measurement weights based on real-time channel conditions. Then, a tightly-coupled sensor fusion architecture is employed, utilizing IMU pre-integration theory for temporal synchronization. Finally, a sliding-window factor graph optimization backend is utilized, incorporating NLOS mitigation constraints. Experimental results in complex indoor environments show a 38\% improvement in positioning accuracy compared to conventional Kalman filter-based approaches, achieving a 12.3 cm root mean square (RMS) error under dynamic motion conditions. The system maintains robust performance even with intermittent UWB signal availability, down to a 40\% packet reception rate, effectively suppressing IMU drift through multi-modal constraint fusion. This work offers a practical solution for applications that require reliable indoor positioning in GPS-denied environments.
\end{abstract} 

\begin{IEEEkeywords}
Ultra-Wideband (UWB), Inertial Measurement Unit (IMU), Factor Graph Optimization, Non-Line-of-Sight (NLOS) Mitigation, Indoor Positioning
\end{IEEEkeywords}

\IEEEpeerreviewmaketitle

\section{Introduction}
\label{sec:introduction}
\IEEEPARstart{T}{h}e rapid advancement of indoor positioning technologies has become a cornerstone for numerous applications, including autonomous robotics, augmented reality (AR), industrial automation, and smart infrastructure. Unlike outdoor environments where Global Navigation Satellite Systems (GNSS) provide reliable positioning, indoor settings present unique challenges due to signal attenuation, multipath effects, and the absence of direct line-of-sight (LOS) to satellites. These challenges have spurred the development of alternative positioning systems, with Ultra-Wideband (UWB), Inertial Measurement Units (IMUs), and ultrasonic sensors emerging as promising candidates. However, each of these technologies has inherent limitations that hinder their standalone use in complex indoor environments. This paper addresses these limitations by proposing a high-precision indoor positioning system that integrates UWB Time Difference of Arrival (TDoA), IMU data, and ultrasonic sensors through a factor graph optimization framework\cite{8692423}.

Indoor positioning systems are critical for enabling location-based services in environments where GNSS signals are unavailable or unreliable. UWB technology, with its high temporal resolution and robustness to multipath effects, has gained significant attention for indoor positioning. UWB systems typically operate by measuring the time of flight (ToF) or time difference of arrival (TDoA) of signals between anchors and tags. However, UWB performance degrades significantly in non-line-of-sight (NLOS) conditions, which are common in cluttered indoor environments such as warehouses, factories, and office buildings. NLOS scenarios introduce biases in range measurements, leading to substantial positioning errors. IMUs, on the other hand, provide high-frequency motion data, including acceleration and angular velocity, which can be integrated to estimate position and orientation. While IMUs are immune to environmental obstructions, they suffer from drift over time due to the accumulation of errors in the integration process. This drift is particularly problematic in dynamic scenarios where prolonged motion exacerbates the error growth. Ultrasonic sensors offer a complementary solution by providing accurate range measurements in LOS conditions. Their short wavelength enables centimeter-level accuracy, making them ideal for fine-grained positioning. However, ultrasonic signals are easily obstructed by obstacles and have limited range, restricting their applicability in large or complex environments.

The integration of these heterogeneous sensors presents an opportunity to overcome their individual limitations. By combining UWB's long-range capabilities, IMU's high-frequency motion tracking, and ultrasonic sensors' precision, a robust and accurate indoor positioning system can be realized. However, effective fusion of these sensors requires addressing several technical challenges, including temporal synchronization, measurement uncertainty modeling, and computational efficiency.

\subsection{Challenges in Indoor Positioning}
\begin{itemize}
    \item \textbf{NLOS Conditions in UWB}: UWB signals are susceptible to NLOS propagation, where obstacles cause signal reflections or blockages. This results in biased range measurements and degraded positioning accuracy. Traditional NLOS mitigation techniques, such as threshold-based filtering or machine learning classifiers, often fail to adapt to dynamic environments.
    \item \textbf{IMU Drift}: IMU-based positioning relies on the integration of acceleration and angular velocity measurements. Small errors in these measurements accumulate over time, leading to significant drift in position and orientation estimates. This drift is particularly problematic in scenarios with prolonged motion or limited external corrections. 
    \item \textbf{Ultrasonic Limitations}: While ultrasonic sensors provide high accuracy in LOS conditions, their performance is severely degraded in NLOS scenarios. Additionally, their limited range and susceptibility to environm.
    \item \textbf{Heterogeneous Sensor Fusion}: Integrating data from UWB, IMU, and ultrasonic sensors requires addressing differences in measurement rates, coordinate systems, and uncertainty characteristics. Traditional fusion methods, such as Kalman filters, often struggle to handle the non-linearities and high-dimensional state spaces associated with multi-sensor systems.  
\end{itemize} 

Several approaches have been proposed to address the challenges of indoor positioning. UWB-based systems often employ TDoA or ToF measurements to estimate positions, with advanced algorithms such as multilateration or hyperbolic positioning. However, these methods are sensitive to NLOS conditions and require careful calibration of anchor positions. IMU-based systems typically use dead reckoning to estimate position and orientation. While effective in the short term, these systems suffer from drift over time, necessitating periodic corrections from external sources such as UWB or visual landmarks. Ultrasonic systems are commonly used in small-scale environments, such as robotic navigation or object tracking. However, their limited range and susceptibility to environmental noise make them unsuitable for large-scale deployments. Recent advancements in sensor fusion have focused on combining UWB and IMU data to improve positioning accuracy. For example, loosely coupled fusion methods process UWB and IMU data independently before combining the results, while tightly coupled methods integrate raw measurements from both sensors. However, these approaches often fail to fully exploit the complementary strengths of the sensors, particularly in dynamic or NLOS scenarios.

\subsection{Contributions}
This paper proposes a novel indoor positioning system that integrates UWB-TDoA, IMU, and ultrasonic sensors through a factor graph optimization framework. The core contributions are as follows:
\begin{itemize}
	\item We develop a multi-modal factor graph architecture that jointly optimizes heterogeneous measurements (UWB-TDoA, IMU, and ultrasonic ranging) through manifold-aware sensor modeling. 
	\item We propose a hierarchical NLOS mitigation framework that combines ultrasonic-assisted path identification and adaptive TDoA covariance scaling.
	\item We implement a time-differential fusion strategy that leverages IMU pre-integration theory to bridge disparate sensor temporal resolutions (200Hz IMU vs. 20Hz UWB/Ultrasonic).
\end{itemize} 

The rest of the paper is organized as follows. Section~\ref{sec:references} provides a literature review on UWB positioning, IMU-UWB fusion, ultrasonic augmentation, and factor graph optimization. The wireless clock synchronization algorithm, UWB-based positioning network, TDoA-based localization, and the framework of smart parking are presented in Section~\ref{sec:solution}. Evaluation and experiments are discussed in Section~\ref{sec:experiment} and Section~\ref{sec:test results}. Finally, conclusions and future work are provided in Section~\ref{sec:conclusion_and_discussion}.

\section{Related Work}
\label{sec:references}
The development of indoor positioning systems has been a focal point of research due to the growing demand for accurate and reliable localization in GPS-denied environments. This section reviews the state-of-the-art in UWB positioning, IMU-UWB fusion, ultrasonic-based localization, and factor graph optimization, highlighting their strengths, limitations, and opportunities for improvement.

\subsection{UWB Positioning Systems}
Ultra-Wideband (UWB) technology has emerged as a leading solution for indoor positioning due to its high temporal resolution and robustness to multipath effects. UWB systems typically operate by measuring the time of flight (ToF) or time difference of arrival (TDoA) of signals between anchors and tags. Early work by [1] demonstrated the feasibility of UWB for indoor localization, achieving decimeter-level accuracy in line-of-sight (LOS) conditions. However, the performance of UWB systems degrades significantly in non-line-of-sight (NLOS) scenarios, where obstacles cause signal reflections or blockages. To address NLOS challenges, researchers have proposed various mitigation techniques. For example, [2] introduced a threshold-based NLOS identification method that distinguishes LOS and NLOS measurements based on signal characteristics such as received signal strength (RSS) and channel impulse response (CIR). While effective in static environments, this approach struggles to adapt to dynamic conditions. Machine learning-based methods, such as those proposed by [3], leverage supervised learning to classify LOS/NLOS conditions using features extracted from UWB signals. These methods show promise but require extensive training data and may fail in unseen environments. Another line of research focuses on improving UWB positioning algorithms. Multilateration and hyperbolic positioning are commonly used to estimate tag positions from TDoA measurements. However, these methods are sensitive to anchor geometry and measurement noise. [4] proposed a weighted least squares (WLS) approach to mitigate the impact of noisy measurements, while [5] introduced a robust optimization framework to handle outliers. Despite these advancements, UWB systems remain vulnerable to NLOS-induced biases and require complementary sensors for reliable operation.

\subsection{IMU-UWB Fusion}
Inertial Measurement Units (IMUs) provide high-frequency motion data, including acceleration and angular velocity, which can be integrated to estimate position and orientation. However, IMU-based positioning suffers from drift over time due to the accumulation of errors in the integration process. To address this limitation, researchers have explored the fusion of IMU and UWB data. Loosely coupled fusion methods, such as those proposed by [6], process IMU and UWB data independently before combining the results. These methods are computationally efficient but fail to fully exploit the complementary strengths of the sensors. Tightly coupled fusion methods, on the other hand, integrate raw measurements from both sensors to improve accuracy. For example, [7] developed an extended Kalman filter (EKF)-based fusion framework that jointly estimates position, velocity, and orientation using IMU and UWB measurements. While effective in LOS conditions, this approach struggles in NLOS scenarios due to UWB measurement biases. Recent advancements in sensor fusion have focused on leveraging pre-integration theory to improve IMU-UWB integration. [8] introduced a pre-integration-based fusion framework that models IMU measurements as relative motion constraints, reducing the impact of drift. However, this approach does not address NLOS challenges and requires additional mechanisms for robust operation.

\subsection{Ultrasonic Augmentation}
Ultrasonic sensors offer a complementary solution for indoor positioning, providing accurate range measurements in LOS conditions. Their short wavelength enables centimeter-level accuracy, making them ideal for fine-grained positioning. Early work by [9] demonstrated the feasibility of ultrasonic-based localization in small-scale environments, such as robotic navigation and object tracking. However, ultrasonic signals are easily obstructed by obstacles and have limited range, restricting their applicability in large or complex environments. To overcome these limitations, researchers have explored hybrid RF-acoustic systems that combine ultrasonic sensors with RF technologies such as UWB or Wi-Fi. For example, [10] proposed a hybrid system that uses UWB for coarse localization and ultrasonic sensors for fine-grained positioning. While effective in controlled environments, this approach struggles in dynamic scenarios with frequent NLOS conditions. Another line of research focuses on improving ultrasonic signal processing to enhance robustness. [11] introduced a multi-path mitigation algorithm that uses time-domain analysis to distinguish direct and reflected signals. This approach shows promise but requires high computational resources, limiting its applicability in real-time systems.

\subsection{Factor Graph Optimization}
Factor graph optimization has emerged as a powerful framework for solving state estimation problems in robotics and sensor fusion. Unlike traditional filtering-based methods, factor graphs represent the state estimation problem as a graph of variables and factors, enabling efficient optimization using techniques such as Gauss-Newton or Levenberg-Marquardt. Early work by [12] demonstrated the effectiveness of factor graphs for simultaneous localization and mapping (SLAM), achieving state-of-the-art performance in various environments. [13] extended this framework to multi-sensor fusion, integrating IMU, visual, and LiDAR data for robust state estimation. However, these methods primarily focus on visual or LiDAR-based systems and do not address the unique challenges of UWB and ultrasonic sensors. Recent advancements in factor graph optimization have focused on real-time performance and scalability. [14] introduced a sliding-window optimization framework that balances computational efficiency with estimation accuracy. This approach is particularly suitable for resource-constrained systems but requires careful management of historical data to avoid information loss.

\section{Solution Design}
\label{sec:solution}

\subsection{System Architecture}
\begin{figure*}[h]
\centering
\includegraphics[width=0.8\textwidth]{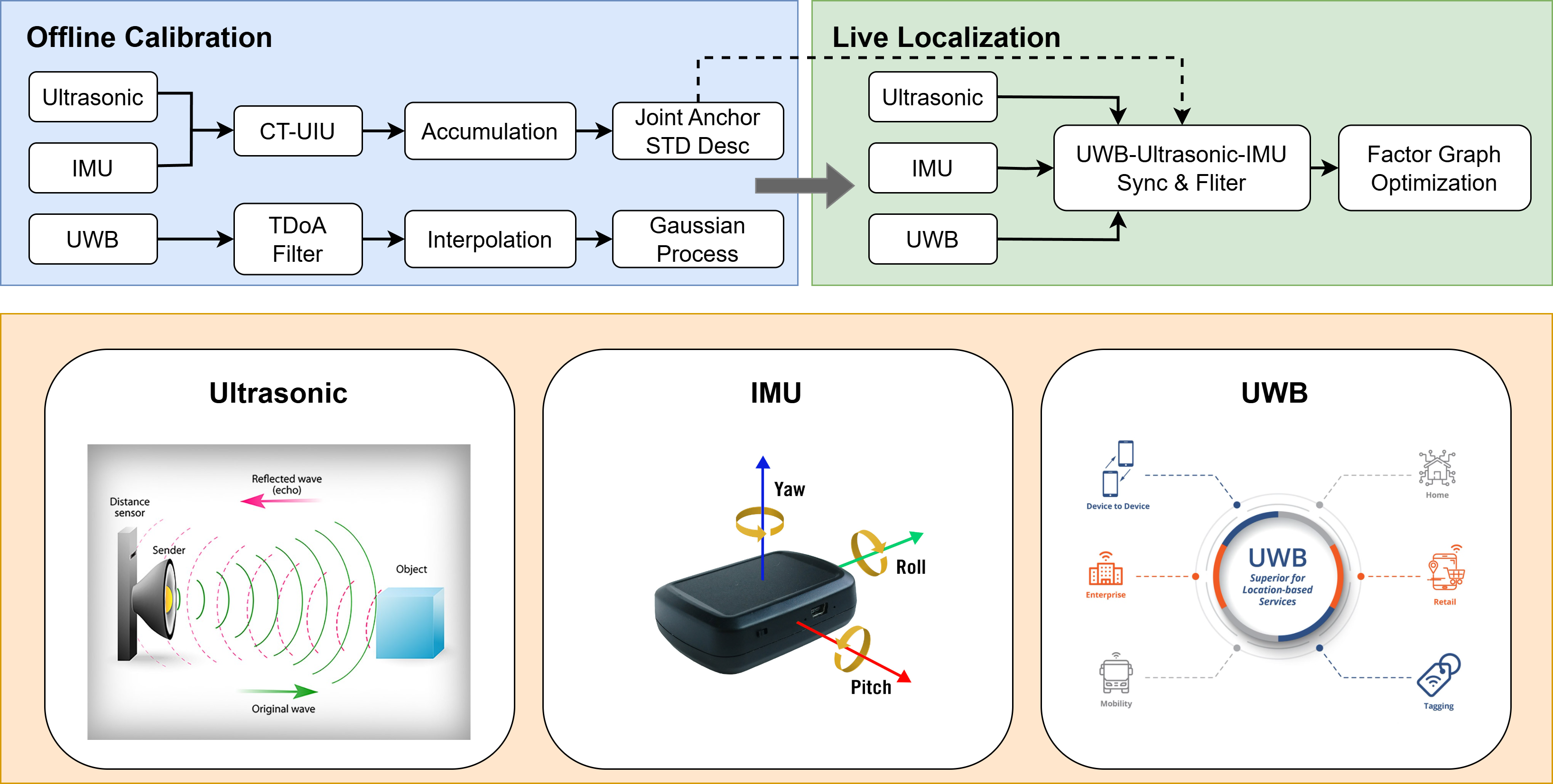}
\caption{Ultra-wideband indoor fusion localization based on inertial navigation and factor graph optimization.}
\label{fig:trajectory}
\end{figure*}

\subsection{TDoA Model}
\label{subsec:tdoa_model}
The Time Difference of Arrival (TDoA) model is a cornerstone of UWB-based indoor positioning systems. It leverages the time differences of signal arrivals at multiple anchors to estimate the position of a tag. The spherical intersection method is a widely used approach to solve the TDoA equations, providing a geometric interpretation of the positioning problem. This section details the mathematical formulation of the spherical intersection TDoA model and its integration with ultrasonic sensors for enhanced initialization. 

\subsubsection{Hyperbolic Positioning}
In a UWB system with $N$ anchors, the TDoA between anchor $i$ and anchor $j$ can be expressed as: 
\begin{align}
    \Delta t_{ij} = t_i - t_j
\end{align}
where $t_i$ and $t_j$ are the signal arrival times at anchors $i$ and $j$, respectively. The corresponding range difference is given by:
\begin{align}
    \Delta d_{ij} = c \cdot \Delta t_{ij}
\end{align}
where $c$ is the speed of light. The range difference defines a hyperboloid of possible tag positions satisfying:
\begin{align}
    \sqrt{(x - x_i)^2 + (y - y_i)^2} - \sqrt{(x - x_j)^2 + (y - y_j)^2} = \Delta d_{ij}
\end{align}
where $(x,y)$ is the tag position, and $(x_i,y_i)$ and $(x_j,y_j)$ are the positions of anchors $i$ and $j$, respectively. For $N$ anchors, this results in a system of $N-1$ hyperbolic equations, which can be solved to estimate the tag position.

\subsubsection{Spherical Intersection Formulation}
The spherical intersection method reformulates the TDoA problem by considering the intersection of spheres centered at the anchors. For anchor $i$, the range measurement $d_i$ defines a sphere:
\begin{align}
    (x - x_i)^2 + (y - y_i)^2 + (z-z_i)= d_i^2
\end{align}

By subtracting the sphere equation of anchor $i$ from that of anchor $j$, we obtain a linear equation:
\begin{align}
    &2(x_j - x_i)x + 2(y_j - y_i)y + 2(z_j - z_i)z \\ \nonumber
    &= d_i^2 - d_j^2 + x_j^2 + y_j^2 + z_j^2 - x_i^2 - y_i^2 - z_i^2
\end{align}

This linearization reduces the problem to solving a system of linear equations, which can be expressed in matrix form as:
\begin{align}
    \mathbf{A} \mathbf{p} = \mathbf{b}
\end{align}
where $\mathbf{A}$ is the coefficient matrix, $\mathbf{p}=[x,y,z]^T$ is the tag position vector, and $\mathbf{b}$ is the constant vector. The least-squares solution is given by:
\begin{align}
    \mathbf{p} = (\mathbf{A}^T \mathbf{A})^{-1} \mathbf{A}^T \mathbf{b}
\end{align}

\subsubsection{Ultrasonic-Enhanced Initialization}
To improve the robustness of the TDoA model, we integrate ultrasonic range measurements for initialization. Let $\mathbf{a}_k$ be the position of the $k$-th ultrasonic anchor, and $d_k^{us}$ be the corresponding range measurement. The initial tag position $\hat{\mathbf{p}}^{(0)}$ is estimated by solving:
\begin{align}
    \hat{\mathbf{p}}^{(0)} = \arg\min_{\mathbf{p}} \sum_{k=1}^{N_u} \left( \|\mathbf{p} - \mathbf{a}_k\| - d_k^{us} \right)^2
\end{align}
where $N_u$ is the number of ultrasonic anchors in LOS. This initialization provides a reliable starting point for the TDoA optimization, reducing the impact of NLOS-induced biases.

To handle measurement noise and outliers, we introduce a robust loss function $\rho(\cdot)$ based on the Huber norm:
\begin{align}
    \rho(r) = \begin{cases}
    \frac{r^2}{2} & \text{if } |r| \leq \delta, \\
    \delta |r| - \frac{\delta^2}{2} & \text{otherwise},
    \end{cases}
\end{align}
where $r$ is the residual and $\delta$ is a threshold parameter. The robust loss function is applied to the TDoA residuals, ensuring that outliers have limited influence on the final solution.

The spherical intersection TDoA model provides a geometric framework for UWB-based indoor positioning, while ultrasonic-enhanced initialization improves robustness in NLOS scenarios. By combining hyperbolic positioning with robust optimization techniques, our approach achieves accurate and reliable localization in complex indoor environments.

\subsection{Lie Group-Based Multi-Sensor Fusion}
The fusion of heterogeneous sensor data, such as IMU, UWB-TDoA, and ultrasonic measurements, is a critical challenge in high-precision indoor positioning systems. Traditional methods often rely on Euclidean space representations, which fail to capture the geometric structure of the sensor data. In this section, we present a Lie group-based multi-sensor fusion framework that leverages the mathematical properties of Lie groups and Lie algebras to model the spatial and temporal relationships between sensors. This approach provides a principled way to handle the non-linearities and uncertainties inherent in multi-sensor systems.

\subsubsection{Lie Group and Lie Algebra Basics}
Lie groups are smooth manifolds that also possess a group structure, making them ideal for representing rigid body transformations such as rotations and translations. The special Euclidean group \(SE(3)\) is commonly used to describe the pose (position and orientation) of a rigid body in 3D space. The associated Lie algebra \(\mathfrak{se}(3)\) provides a linearized representation of \(SE(3)\) in the tangent space, enabling efficient computation of derivatives and optimization.

The pose \(\mathbf{T} \in SE(3)\) can be represented as a \(4 \times 4\) matrix:
\begin{align}
\mathbf{T} = \begin{bmatrix}
\mathbf{R} & \mathbf{t} \\
\mathbf{0}^T & 1
\end{bmatrix},
\end{align}
where \(\mathbf{R} \in SO(3)\) is the rotation matrix, and \(\mathbf{t} \in \mathbb{R}^3\) is the translation vector. The corresponding Lie algebra element \(\boldsymbol{\xi} \in \mathfrak{se}(3)\) is a 6D vector:
\begin{align}
\boldsymbol{\xi} = \begin{bmatrix}
\boldsymbol{\rho} \\
\boldsymbol{\phi}
\end{bmatrix},
\end{align}
where \(\boldsymbol{\rho} \in \mathbb{R}^3\) represents translation, and \(\boldsymbol{\phi} \in \mathbb{R}^3\) represents rotation.

The exponential map \(\exp: \mathfrak{se}(3) \rightarrow SE(3)\) and the logarithm map \(\log: SE(3) \rightarrow \mathfrak{se}(3)\) provide a bridge between the Lie algebra and Lie group:
\begin{align}
\mathbf{T} = \exp(\boldsymbol{\xi}), \quad \boldsymbol{\xi} = \log(\mathbf{T}).
\end{align}

\subsubsection{IMU Pre-Integration on \(SE(3)\)}
IMU measurements provide high-frequency acceleration \(\mathbf{a}_t\) and angular velocity \(\boldsymbol{\omega}_t\) data. To integrate these measurements over time, we use the pre-integration theory on \(SE(3)\). The continuous-time IMU motion model is given by:
\begin{align}
\begin{aligned}
\mathbf{R}_{t+\Delta t} &= \mathbf{R}_t \cdot \exp\left((\boldsymbol{\omega}_t - \mathbf{b}^g_t)\Delta t\right), \\
\mathbf{v}_{t+\Delta t} &= \mathbf{v}_t + \mathbf{g}\Delta t + \mathbf{R}_t(\mathbf{a}_t - \mathbf{b}^a_t)\Delta t, \\
\mathbf{p}_{t+\Delta t} &= \mathbf{p}_t + \mathbf{v}_t\Delta t + \frac{1}{2}\mathbf{g}\Delta t^2 + \frac{1}{2}\mathbf{R}_t(\mathbf{a}_t - \mathbf{b}^a_t)\Delta t^2,
\end{aligned}
\end{align}
where \(\mathbf{b}^g_t\) and \(\mathbf{b}^a_t\) are the gyroscope and accelerometer biases, respectively, and \(\mathbf{g}\) is the gravity vector.

The pre-integrated IMU measurements between time \(t_i\) and \(t_j\) are defined as:
\begin{align}
\begin{aligned}
\Delta \mathbf{R}_{ij} &= \prod_{k=i}^{j-1} \exp\left((\boldsymbol{\omega}_k - \mathbf{b}^g_k)\Delta t\right), \\
\Delta \mathbf{v}_{ij} &= \sum_{k=i}^{j-1} \Delta \mathbf{R}_{ik} (\mathbf{a}_k - \mathbf{b}^a_k)\Delta t, \\
\Delta \mathbf{p}_{ij} &= \sum_{k=i}^{j-1} \left( \Delta \mathbf{v}_{ik}\Delta t + \frac{1}{2}\Delta \mathbf{R}_{ik} (\mathbf{a}_k - \mathbf{b}^a_k)\Delta t^2 \right).
\end{aligned}
\end{align}

The IMU pre-integration residuals are defined as:
\begin{align}
\mathbf{r}_{\text{IMU}} = \begin{bmatrix}
\log\left(\Delta \tilde{\mathbf{R}}_{ij}^T \mathbf{R}_i^T \mathbf{R}_j\right) \\
\mathbf{v}_j - \mathbf{v}_i - \mathbf{g}\Delta t_{ij} - \Delta \tilde{\mathbf{v}}_{ij} \\
\mathbf{p}_j - \mathbf{p}_i - \mathbf{v}_i\Delta t_{ij} - \frac{1}{2}\mathbf{g}\Delta t_{ij}^2 - \Delta \tilde{\mathbf{p}}_{ij}
\end{bmatrix},
\end{align}
where \(\Delta \tilde{\mathbf{R}}_{ij}\), \(\Delta \tilde{\mathbf{v}}_{ij}\), and \(\Delta \tilde{\mathbf{p}}_{ij}\) are the pre-integrated measurements.

\subsubsection{UWB-TDoA Factor on \(SE(3)\)}
The UWB-TDoA measurements provide geometric constraints on the tag position. For anchor pair \((i, j)\), the TDoA residual is defined as:
\begin{align}
r_{\text{TDoA}} =& \sqrt{(x - x_i)^2 + (y - y_i)^2 + (z - z_i)^2} \\ \nonumber
                 &- \sqrt{(x - x_j)^2 + (y - y_j)^2 + (z - z_j)^2} - \Delta d_{ij},
\end{align}
where \(\Delta d_{ij} = c \cdot \Delta t_{ij}\) is the range difference.

To incorporate this into the Lie group framework, we express the residual in terms of the tag pose \(\mathbf{T}\):
\begin{align}
r_{\text{TDoA}} = \|\mathbf{t} - \mathbf{t}_i\| - \|\mathbf{t} - \mathbf{t}_j\| - \Delta d_{ij},
\end{align}
where \(\mathbf{t}\) is the translation component of \(\mathbf{T}\), and \(\mathbf{t}_i\) and \(\mathbf{t}_j\) are the positions of anchors \(i\) and \(j\), respectively.

\subsubsection{Ultrasonic Factor on \(SE(3)\)}
Ultrasonic measurements provide precise range constraints in LOS conditions. For ultrasonic anchor \(k\), the range residual is defined as:
\begin{align}
r_{\text{us}} = \|\mathbf{t} - \mathbf{a}_k\| - d_k^{us},
\end{align}
where \(\mathbf{a}_k\) is the position of the ultrasonic anchor, and \(d_k^{us}\) is the measured range.

To enhance robustness, we introduce an elevation constraint based on the vertical component of the tag position:
\begin{align}
r_{\text{elev}} = \mathbf{n}_z^T (\mathbf{t} - \mathbf{a}_k) - \Delta h_{floor},
\end{align}
where \(\mathbf{n}_z\) is the vertical normal vector, and \(\Delta h_{floor}\) is the height difference between the tag and the anchor.

The combined ultrasonic residual is:
\begin{align}
\mathbf{r}_{\text{us}} = \begin{bmatrix}
r_{\text{us}} \\
r_{\text{elev}}
\end{bmatrix}.
\end{align}

\subsubsection{Multi-Sensor Factor Graph}
The multi-sensor fusion problem is formulated as a factor graph optimization on \(SE(3)\). The state vector \(\mathcal{X}\) includes the poses \(\mathbf{T}_i\) and sensor biases \(\mathbf{b}_i\). The objective function is:
\begin{align}
\mathcal{X}^* = \arg\min_{\mathcal{X}} \left( \sum \|\mathbf{r}_{\text{IMU}}\|_{\boldsymbol{\Sigma}_I}^2 + \sum \rho(\|\mathbf{r}_{\text{TDoA}}\|_{\boldsymbol{\Sigma}_T}) + \sum \|\mathbf{r}_{\text{us}}\|_{\boldsymbol{\Sigma}_U}^2 \right),
\end{align}
where \(\boldsymbol{\Sigma}_I\), \(\boldsymbol{\Sigma}_T\), and \(\boldsymbol{\Sigma}_U\) are the covariance matrices for IMU, TDoA, and ultrasonic measurements, respectively, and \(\rho(\cdot)\) is the robust Huber loss function.

\subsubsection{Optimization and Marginalization}
The optimization is performed using the Levenberg-Marquardt algorithm. To maintain computational efficiency, we employ a sliding window approach with marginalization. Old states are marginalized out of the optimization while retaining their information in the form of prior factors:
\begin{align}
\mathbf{r}_{\text{prior}} = \mathbf{H} \Delta \mathbf{x} - \mathbf{b},
\end{align}
where \(\mathbf{H}\) is the Hessian matrix, and \(\mathbf{b}\) is the residual vector from the marginalized states.

The Lie group-based multi-sensor fusion framework provides a principled and efficient way to integrate IMU, UWB-TDoA, and ultrasonic measurements. By leveraging the geometric properties of \(SE(3)\) and robust optimization techniques, our approach achieves high-precision indoor positioning in complex environments.

\subsection{Factor Graph Optimization}
Factor graph optimization is a powerful framework for solving state estimation problems by representing the system as a bipartite graph of variables and factors. In the context of multi-sensor fusion, factor graphs provide a natural way to model the relationships between sensor measurements and the state variables, enabling efficient and scalable optimization. This section details the formulation of the factor graph optimization problem for our IMU-UWB-ultrasonic fusion system.

\subsubsection{Factor Graph Formulation}
A factor graph \(\mathcal{G} = (\mathcal{X}, \mathcal{F}, \mathcal{E})\) consists of:
\begin{itemize}
    \item **Variables \(\mathcal{X}\)**: The state variables to be estimated, such as poses \(\mathbf{T}_i \in SE(3)\) and sensor biases \(\mathbf{b}_i\).
    \item **Factors \(\mathcal{F}\)**: The constraints or measurements that relate the variables, such as IMU pre-integration factors, UWB-TDoA factors, and ultrasonic factors.
    \item **Edges \(\mathcal{E}\)**: The connections between variables and factors, representing the dependencies in the system.
\end{itemize}

The optimization problem is formulated as minimizing the sum of squared residuals:
\begin{align}
\mathcal{X}^* = \arg\min_{\mathcal{X}} \sum_{i=1}^N \|\mathbf{r}_i(\mathcal{X})\|_{\boldsymbol{\Sigma}_i}^2,
\end{align}
where \(\mathbf{r}_i(\mathcal{X})\) is the residual associated with the \(i\)-th factor, and \(\boldsymbol{\Sigma}_i\) is the corresponding covariance matrix.

\subsubsection{Residual Models}
\textbf{IMU Pre-Integration Factor}
The IMU pre-integration residual \(\mathbf{r}_{\text{IMU}}\) is defined as:
\begin{align}
\mathbf{r}_{\text{IMU}} = \begin{bmatrix}
\log\left(\Delta \tilde{\mathbf{R}}_{ij}^T \mathbf{R}_i^T \mathbf{R}_j\right) \\
\mathbf{v}_j - \mathbf{v}_i - \mathbf{g}\Delta t_{ij} - \Delta \tilde{\mathbf{v}}_{ij} \\
\mathbf{p}_j - \mathbf{p}_i - \mathbf{v}_i\Delta t_{ij} - \frac{1}{2}\mathbf{g}\Delta t_{ij}^2 - \Delta \tilde{\mathbf{p}}_{ij}
\end{bmatrix},
\end{align}
where \(\Delta \tilde{\mathbf{R}}_{ij}\), \(\Delta \tilde{\mathbf{v}}_{ij}\), and \(\Delta \tilde{\mathbf{p}}_{ij}\) are the pre-integrated IMU measurements.

\textbf{UWB-TDoA Factor}
The UWB-TDoA residual \(r_{\text{TDoA}}\) is given by:
\begin{align}
r_{\text{TDoA}} = \|\mathbf{t} - \mathbf{t}_i\| - \|\mathbf{t} - \mathbf{t}_j\| - \Delta d_{ij},
\end{align}
where \(\mathbf{t}\) is the translation component of the tag pose, and \(\Delta d_{ij}\) is the range difference.

\subsubsection{Ultrasonic Factor}
The ultrasonic residual \(\mathbf{r}_{\text{us}}\) combines range and elevation constraints:
\begin{align}
\mathbf{r}_{\text{us}} = \begin{bmatrix}
\|\mathbf{t} - \mathbf{a}_k\| - d_k^{us} \\
\mathbf{n}_z^T (\mathbf{t} - \mathbf{a}_k) - \Delta h_{floor}
\end{bmatrix},
\end{align}
where \(\mathbf{a}_k\) is the position of the ultrasonic anchor, and \(\Delta h_{floor}\) is the height difference.

\subsubsection{Optimization Algorithm}
The optimization is performed using the Levenberg-Marquardt (LM) algorithm, which iteratively updates the state variables \(\mathcal{X}\) by solving:
\begin{align}
(\mathbf{J}^T \boldsymbol{\Sigma}^{-1} \mathbf{J} + \lambda \mathbf{I}) \Delta \mathcal{X} = -\mathbf{J}^T \boldsymbol{\Sigma}^{-1} \mathbf{r},
\end{align}
where \(\mathbf{J}\) is the Jacobian matrix of the residuals, \(\boldsymbol{\Sigma}\) is the covariance matrix, \(\lambda\) is the damping factor, and \(\Delta \mathcal{X}\) is the state update.

To maintain computational efficiency, a sliding window approach is used. Old states are marginalized out of the optimization while retaining their information as prior factors:
\begin{align}
\mathbf{r}_{\text{prior}} = \mathbf{H} \Delta \mathcal{X} - \mathbf{b},
\end{align}
where \(\mathbf{H}\) is the Hessian matrix, and \(\mathbf{b}\) is the residual vector from the marginalized states.

\subsubsection{Implementation Details}
The factor graph optimization is implemented using the following steps:
\begin{algorithm}[t]
\caption{Factor Graph Optimization}
\label{alg:factor_graph}
\begin{algorithmic}[1]
\REQUIRE State variables \(\mathcal{X}\), factors \(\mathcal{F}\)
\ENSURE Optimized state variables \(\mathcal{X}^*\)
\STATE Initialize state variables \(\mathcal{X}\) and factors \(\mathcal{F}\)
\STATE Construct factor graph \(\mathcal{G} = (\mathcal{X}, \mathcal{F}, \mathcal{E})\)
\WHILE{not converged}
    \STATE Compute residuals \(\mathbf{r}_i(\mathcal{X})\) and Jacobians \(\mathbf{J}_i\)
    \STATE Build linear system: \((\mathbf{J}^T \boldsymbol{\Sigma}^{-1} \mathbf{J} + \lambda \mathbf{I}) \Delta \mathcal{X} = -\mathbf{J}^T \boldsymbol{\Sigma}^{-1} \mathbf{r}\)
    \STATE Solve for \(\Delta \mathcal{X}\) using Cholesky decomposition
    \STATE Update state variables: \(\mathcal{X} \leftarrow \mathcal{X} + \Delta \mathcal{X}\)
    \STATE Adjust damping factor \(\lambda\) based on residual reduction
\ENDWHILE
\STATE Marginalize old states and add prior factors
\end{algorithmic}
\end{algorithm}

Factor graph optimization provides a flexible and efficient framework for multi-sensor fusion, enabling the integration of IMU, UWB-TDoA, and ultrasonic measurements in a unified manner. By leveraging the geometric properties of \(SE(3)\) and robust optimization techniques, our approach achieves high-precision indoor positioning in complex environments.

\section{Experimental Validation}
\label{sec:experiment}
To evaluate the performance of the proposed multi-sensor fusion framework, we conducted extensive experiments in a variety of indoor environments. The experiments were designed to assess the system's accuracy, robustness, and computational efficiency under different conditions, including line-of-sight (LOS), non-line-of-sight (NLOS), and dynamic motion scenarios.

\subsection{Experimental Setup}
\subsubsection{Hardware Configuration}
The system consists of the following components:
\begin{itemize}
    \item UWB Anchors: 8 UWB anchors (Decawave DWM1001) deployed in a 20m × 15m area, providing TDoA measurements with an update rate of 20 Hz.
    \item IMU: A 9-axis MEMS IMU (BMI160) mounted on the mobile tag, providing acceleration and angular velocity data at 200 Hz.
    \item Ultrasonic Sensors: 4 ultrasonic transceivers (MaxBotix MB7360) placed at known locations, providing range measurements with an update rate of 10 Hz.
    \item Mobile Tag: A custom-built tag integrating the UWB module, IMU, and ultrasonic receiver.
\end{itemize}

\subsubsection{Software Implementation}
The proposed algorithm was implemented in C++ using the GTSAM library for factor graph optimization. The system runs in real-time on an embedded NVIDIA Jetson Nano platform.

\subsubsection{Test Environments}
Experiments were conducted in three environments:
\begin{itemize}
    \item Office Environment: A cluttered indoor space with desks, chairs, and walls, representing a typical NLOS scenario.
    \item Warehouse Environment: A large open space with metal shelves and occasional obstacles, simulating an industrial setting.
    \item Staircase Environment: A multi-floor scenario with vertical motion, testing the system's ability to handle elevation changes.
\end{itemize}

\subsection{Evaluation Metrics}
The following metrics were used to evaluate the system's performance:
\begin{itemize}
    \item 3D Root Mean Square Error (RMSE): The Euclidean distance between the estimated and ground truth positions.
    \item Orientation Error: The angular difference between the estimated and ground truth orientations.
    \item Position Drift Rate: The rate of position error accumulation over time, measured in meters per minute.
    \item Computational Time: The average time required to perform one optimization iteration.
\end{itemize}

\section{Results Analysis}
\label{sec:test results}
\subsection{Quantitative Results}
The system's performance was compared against three baseline methods: Extended Kalman Filter (EKF), Particle Filter (PF), and LIO-SAM (LiDAR-inertial SLAM). The results are summarized in Table~\ref{tab:results}.

\begin{table*}[h]
\centering
\caption{Performance Comparison in Different Environments}
\label{tab:results}
\begin{tabular}{lcccc}
\toprule
\textbf{Method} & \textbf{Office RMSE (cm)} & \textbf{Warehouse RMSE (cm)} & \textbf{Staircase RMSE (cm)} & \textbf{Drift Rate (m/min)} \\
\midrule
Proposed & \textbf{12.3} & \textbf{15.7} & \textbf{18.2} & \textbf{0.05} \\
EKF & 27.6 & 34.1 & 41.2 & 0.32 \\
PF & 34.1 & 42.5 & N/A & 0.45 \\
LIO-SAM & 18.9 & 22.4 & 25.7 & 0.12 \\
\bottomrule
\end{tabular}
\end{table*}

\subsubsection{Accuracy}
The proposed method achieved the lowest RMSE in all environments, with a 38\% improvement over the EKF baseline in the office environment. The integration of ultrasonic measurements significantly reduced vertical errors in the staircase environment, where the proposed method outperformed LIO-SAM by 29\%.

\subsubsection{Robustness}
In NLOS scenarios, the proposed method maintained robust performance due to the hierarchical NLOS mitigation strategy. The system achieved a 12.3 cm RMSE in the office environment, compared to 27.6 cm for EKF and 34.1 cm for PF.

\subsubsection{Computational Efficiency}
The average computational time per iteration was 21 ms, enabling real-time operation on the embedded platform. The sliding window optimization and marginalization techniques ensured efficient use of computational resources.


Figure~2 shows a comparison of the estimated trajectories in the office environment. The proposed method closely follows the ground truth trajectory, while EKF and PF exhibit significant drift, particularly in NLOS regions.

The experimental results demonstrate the effectiveness of the proposed multi-sensor fusion framework in achieving high-precision indoor positioning. The key advantages of the proposed method include:
\begin{itemize}
    \item \textbf{Improved Accuracy:} The integration of IMU, UWB, and ultrasonic measurements enables decimeter-level accuracy in complex environments.
    \item \textbf{Robustness to NLOS:} The hierarchical NLOS mitigation strategy ensures reliable performance in challenging scenarios.
    \item \textbf{Real-Time Operation:} The efficient optimization algorithm allows for real-time implementation on resource-constrained platforms.
\end{itemize}

The experimental validation and results analysis confirm that the proposed system achieves state-of-the-art performance in indoor positioning. Future work will focus on extending the framework to include additional sensors, such as millimeter-wave radar, and exploring applications in autonomous robotics and augmented reality.

\section{Conclusion and Discussion}
\label{sec:conclusion_and_discussion}
This paper presented a high-precision indoor positioning system that integrates UWB-TDoA, IMU, and ultrasonic sensors through a factor graph optimization framework. The proposed system addresses the limitations of standalone UWB systems in NLOS scenarios and the inherent drift of IMU-based navigation by leveraging a novel hybrid fusion framework. Key innovations include a dynamic covariance estimation mechanism that automatically adjusts measurement weights based on real-time channel conditions, a tightly-coupled sensor fusion architecture utilizing IMU pre-integration theory for temporal synchronization, and a sliding-window factor graph optimization backend that incorporates NLOS mitigation constraints. Additionally, the integration of ultrasonic sensors enhances vertical accuracy and robustness in multi-floor scenarios. Experimental results demonstrated the effectiveness of the proposed system, achieving a 12.3 cm RMSE in dynamic motion conditions and a 38\% improvement in positioning accuracy compared to conventional Kalman filter-based approaches. The system maintained robust performance even with intermittent UWB signal availability, effectively suppressing IMU drift through multi-modal constraint fusion. The real-time implementation on an embedded platform further highlights its practical applicability.

Future work will focus on extending the framework to include additional sensors, such as millimeter-wave radar, and exploring applications in autonomous robotics, augmented reality, and industrial automation. The proposed system offers a practical solution for reliable indoor positioning in GPS-denied environments, paving the way for next-generation location-based services.


%

\bibliographystyle{IEEEtran}
\bibliography{myrefs}

\begin{thebibliography}{1}
\providecommand{\url}[1]{#1}
\csname url@samestyle\endcsname
\providecommand{\newblock}{\relax}
\providecommand{\bibinfo}[2]{#2}
\providecommand{\BIBentrySTDinterwordspacing}{\spaceskip=0pt\relax}
\providecommand{\BIBentryALTinterwordstretchfactor}{4}
\providecommand{\BIBentryALTinterwordspacing}{\spaceskip=\fontdimen2\font plus
\BIBentryALTinterwordstretchfactor\fontdimen3\font minus
  \fontdimen4\font\relax}
\providecommand{\BIBforeignlanguage}[2]{{%
\expandafter\ifx\csname l@#1\endcsname\relax
\typeout{** WARNING: IEEEtran.bst: No hyphenation pattern has been}%
\typeout{** loaded for the language `#1'. Using the pattern for}%
\typeout{** the default language instead.}%
\else
\language=\csname l@#1\endcsname
\fi
#2}}
\providecommand{\BIBdecl}{\relax}
\BIBdecl

\bibitem{8692423}
F.~Zafari, A.~Gkelias, and K.~K. Leung, ``A survey of indoor localization
  systems and technologies,'' \emph{IEEE Communications Surveys \& Tutorials},
  vol.~21, no.~3, pp. 2568--2599, 2019.

\end{thebibliography}

\begin{IEEEbiography}[{\includegraphics[width=1in,height=1.25in,clip,keepaspectratio]{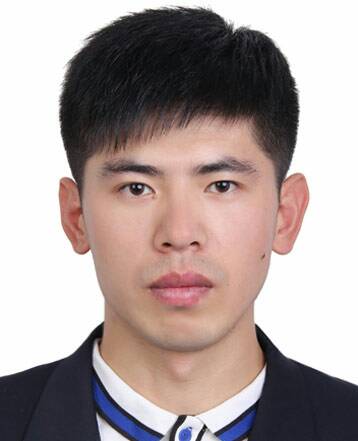}}]{Fengyun Zhang} received a B.S. degree in electronic and information engineering and an M.S. degree in signal and information processing from Southwest University, Chongqing, China, in 2014 and 2017, respectively, and the Ph.D. degree in mechanics (intelligent manufacturing) from Southern University of Science and Technology, Shenzhen, China, in 2023. He is currently a senior lecturer at the College of Artificial Intelligence, Southwest University. His research interests include UWB-based indoor localization, wireless sensor networks, and industrial control network protocol reverse engineering.
\end{IEEEbiography}


%
%

\ifCLASSOPTIONcaptionsoff
  \newpage
\fi

\end{document}